\documentclass[twocolumn,showpacs,prl,superscriptaddress]{revtex4}

\usepackage{graphicx,amsmath,txfonts}

\begin{document}
\title{Emergent Hierarchical Structures in Multiadaptive Games}

\author{Sungmin Lee}
\affiliation{IceLab, Department of Physics, Ume{\aa} University, 90187 Ume\aa, Sweden}
\author{Petter Holme}
\affiliation{IceLab, Department of Physics, Ume{\aa} University, 90187 Ume\aa, Sweden}
\affiliation{Department of Energy Science, Sungkyunkwan University, Suwon 440--746, Korea}
\author{Zhi-Xi Wu}
\affiliation{Institute of Computational Physics and Complex Systems, Lanzhou University, Lanzhou, Gansu 730000, China}
\affiliation{IceLab, Department of Physics, Ume{\aa} University, 90187 Ume\aa, Sweden}

\begin{abstract}
We investigate a game-theoretic model of a social system where both the rules of the game and the interaction structure are shaped by the behavior of the agents. We call this type of model, with several types of feedback couplings from the behavior of the agents to their environment, a multiadaptive game. Our model has a complex behavior with several regimes of different dynamic behavior accompanied by different network topological properties. Some of these regimes are characterized by heterogeneous, hierarchical interaction networks, where cooperation and network topology coemerge from the dynamics.
\end{abstract}
\maketitle


Game theory is a language for describing systems in
biology, economy, and society where the success of an
agent depends both on its own behavior and the behaviors
of others. Perhaps the most important question for gametheoretic
research is to map out the conditions for cooperation
to emerge among egoistic individuals~\cite{axe:evo_maynard_nowak:evodyn}. To this
end, researchers have developed a number of different
types of models, capturing different game-theoretic scenarios.
In this Letter, we investigate a generalization in a
new direction, relaxing constraints of other models, with
feedback effects at different levels to the behavior of the
agents.

In most game-theoretical studies, the rules of the game
are fixed in time, but in real systems there is a feedback
from the behavior of the agents to the environment and thus
to the rules of the game. The payoff of a playerÕs action in a
specific situation is parametrized by payoff matrices. A
straightforward way of modeling feedback from the system
to the rules is to let the entries of the matrices be variables,
dependent on the state of the system~\cite{tomochi}. Another feature
that often is modeled as static when, in reality, it does not
have to be, is the contact structure. If agents can change
their interaction patterns in response to the outcome of the
game, then the model will also capture the social network
dynamics. Such adaptive-network models~\cite{egui:evo_ptn:coevo,wli2007,zimmermann2000} can address
a wide range of problems: not only how interaction
determines the evolution of cooperation, but also how the
interaction patterns themselves emerge. In this Letter we
investigate a situation where agents can adjust their social
ties to maximize their payoffs and the collective behavior
of the agents shapes the rules of the game. Our model is an
adaptive-network model with adaptive payoff matrices---a
multiadaptive game, for short.

A classic model for studying the evolution of cooperation
in spatial game theory is the Nowak-May (NM) game~\cite{nowakmay} (technically speaking, on the border between the archetypical
prisonerÕs dilemma and chicken games). It captures
a situation where at any moment defection has the highest
expected payoff, but under some conditions agents can do
better in a long time perspective by establishing trust and
cooperation. An interaction in the NM game gives the
following payoff: zero to anyone interacting with a defector
($D$), one to a cooperator ($C$) meeting another cooperator,
and $b>1$ to a $D$ meeting a $C$. This model has been
used to explain the emergence of cooperation among egoistic
agents in disciplines as diverse as political science,
economics, and biology~\cite{axe:evo_maynard_nowak:evodyn}, and will be the starting point of our work.

In our model we place $L\times L$ (in this Letter, we use $L=100$) agents on a square grid with fixed boundary condition. Besides interacting with $n$ local spatial neighbors ($n=4$, $3$, and $2$ for internal, boundary, and corner agents, respectively), each agent has one additional link free to optimize its position in the interaction network~\cite{wli2007}. The rationale behind this arrangement is that people invest more in their spatially close contacts (e.g., family and coworkers) and thus are less likely  to break these, whereas the long-range edges are more businesslike and open to optimization. In sociology this situation goes by the name ``strength of weak ties''~\cite{grano:weak}.

\begin{figure}[h]
\includegraphics[width=\linewidth]{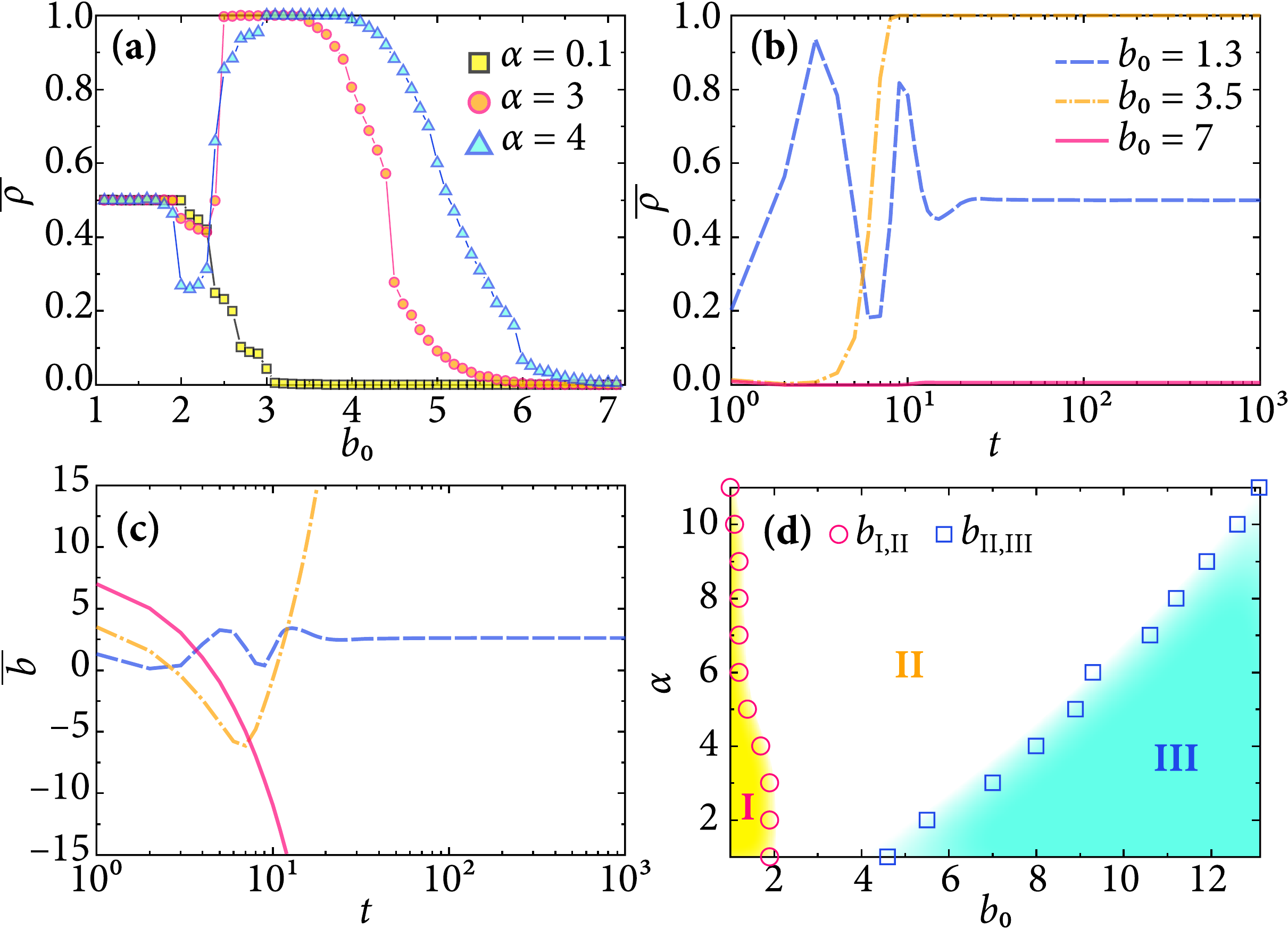}
\caption{(Color online) Parameter dependence of the game reflected in the temptation and average cooperator density.  (a) shows average density of cooperators, $\bar\rho $, as a function of the initial temptation, $b_0$, with $\alpha=0.1$, $3$ and $4$. The bar represents points averaged over the last $500$ (of $10^3$) steps. (b) and (c) correspond to the time evolution of $\bar\rho $ and $\bar b$, respectively, for  different values of $b_0$. (d) shows the diagram over the three regions in $\alpha$-$b_0$ space.  The curves are averages over $10^4$ runs.
}\label{rhocall}
\end{figure}

In the NM game there is one parameter, the temptation
to defect b, representing external conditions of the game
(``society'' in a social interpretation of the game, ``environment''
in the context of evolutionary biology). In this work,
we investigate the case when $b$ is higher in a uniformly rich
society, whereas the motivation to cooperate is higher in a
society in unrest. Assuming a linear dependence of the
temptation to defect on the prosperity, in our case measured
by the density $\rho$ of cooperators in the population, we
use a response function~\cite{tomochi}
\begin{equation}\label{tupdate}
b(t+1)=b(t)+\alpha[\rho (t)-\rho^*],
\end{equation}
where we choose $\rho^*$ (representing a neutral cooperation
level from the societyÕs perspective) as $1/2$ for simplicity.
The value of $\alpha$ controls the strength of feedback from the
environment to the game rules. We will, unless otherwise stated, use $\alpha=4$. We update the state of the system, both strategies and long-range linked neighbors of the agents, synchronously. At a time step, each agent $i$ acquires payoff $u_i$ by playing the Nowak-May game with all its local and long-range neighbors. When an agent $i$, updating its strategy, has a higher payoff than its neighbors, nothing happens. Otherwise, $i$ adopts the strategy of the neighbor $j$ with the highest payoff with a probability $\Pi(i\rightarrow j)$, and simultaneously rewires its free link to the long-range neighbor of $j$. Following Ref.~\cite{wli2007}, we use
\begin{equation}\label{rewiringprob}
\Pi(i\rightarrow j)=1\big /\{1+\exp[-\beta (u_j-u_i)]\}~,
\end{equation}
where $\beta$ controls the noise in the choice of whom to
imitate. This way of parametrizing noise is further discussed
in Ref. ~\cite{szabotoke}. We use $\beta=1$ in our present study, which is enough to create heterogeneous structures but not enough to overshadow the strategies as a factor in the dynamics.

\begin{figure*}
\includegraphics[width=\linewidth]{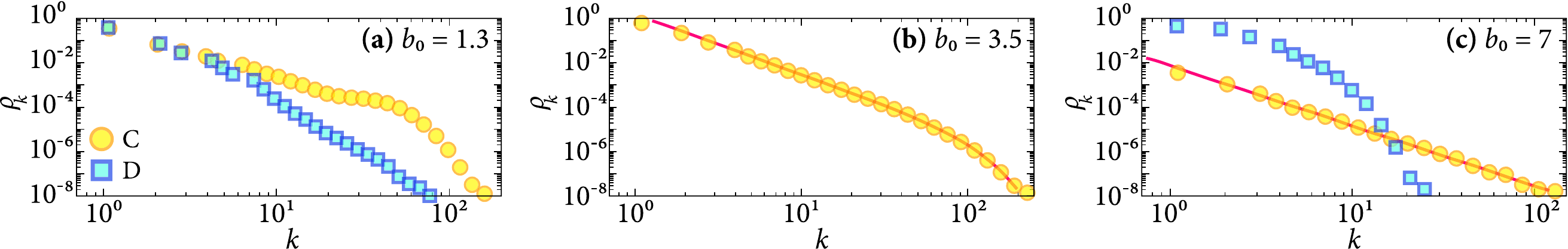}
\caption{(Color online) Correlations between the strategy and network structure. Circles (squares) correspond to the average density of cooperators (defectors) with degree $k$, $\rho_k$. (a) is for $b_0=1.3$ (region I), (b) is for $b_0=3.5$ (region II), and (c) is for $b_0=7$ (region III). In panels (b) and (c), the exponent of the power-law is $2.7\pm 0.1$.}\label{rhocdk}
\end{figure*}

Turning to the numerical results, in Fig.~\ref{rhocall}(a), we plot the average density of cooperators $\bar\rho$ as a function of $b_0$ for three values of $\alpha$. For example, if $\alpha=4$ and $b_0\lesssim 2$, the system converge with certainty to a state with $\bar\rho\approx\rho^*=1/2$. We call the region of parameter space with this behavior \textit{region I} and denote the large-$b_0$ border of this region $b_{\mathrm{I,II}}$. For the $\alpha$-values of Fig.~\ref{rhocall}(a) $b_{\mathrm{I,II}}\approx 2$. For $b_0\gtrsim b_{\mathrm{II,III}}$ cooperation vanishes. We call this part of $\alpha$-$b_0$ space region III.  Between these extremes, there is a region II of complex behavior where, depending on $b_0$, the cooperation density converges to $1$, $\rho^*$ (at least a value very close to $\rho^*$) or $0$ with probabilities depending on $\alpha$ and $b_0$. With increasing $b_0$, the probability that the system end in all-$C$ decreases, and vanishes completely at $b_{\mathrm{II,III}}$.

In Figs.~\ref{rhocall}(b) and (c) we display trajectories of $\rho$ and $b$,
averaged over $10^4$ runs, for different $b_0$ values. These
curves show the system stabilizing to a steady cooperation
level after about $50$ time steps. These transient oscillations
can be explained by the adaptive payoff dynamics.
Assuming a well-mixed case, in which the strategy adoption
rate is proportional to the relative success of the
strategies, one can approximate the dynamics by the replicator
equation system
\begin{subequations}
\begin{eqnarray}
\frac{\mathrm{d}\rho}{\mathrm{d}t} &=&\left\{\begin{array}{ll}\rho^2(1-\rho)(1-b) &\mbox{if~} \rho\in[0,1]\\0 & \mbox{otherwise}\end{array}\right.\label{eq:rho_rep_eq}\\
\frac{\mathrm{d}b}{\mathrm{d}t} &=&\alpha(\rho-\rho^*) .
\end{eqnarray}
\end{subequations}
The factors $\rho^2$ and $1-\rho$ of Eq.~(\ref{eq:rho_rep_eq}) give the fixed points $\rho=0$ and $1$. From these equations, we can also understand the oscillatory behavior of Fig.~\ref{tupdate}(b) and (c).   If $b>1$
and $\rho>\rho^*$, then $b$ will increase and $\rho$ decrease. This will,
after some time, make $\rho<\rho^*$ and thus $db/dt$ negative. If
$db/dt$ is negative, then $d\rho/dt$ will eventually become
positive. Taken together, this explains the cyclic behavior.
Such oscillations---growing and shrinking $C$ ($D$) clusters
that drive the oscillations in $\rho$---can be seen with our Java
applet of the model~\cite{java}. For all parameter values we study,
the cyclic behavior will either increase in amplitude until $\rho$
reaches a fixed point, or be dampened to the fixed point
close to $\rho^*$. The perhaps most interesting observation is the
onset of the all-$C$ state. As an example, for $b_0 = 3.5$ in
Fig.\ref{tupdate}(b), $\bar\rho$ starts increasing again, but it is too late---the emergence of a $C$ hub, combined with the fact that $b$ is
still smaller than $1$, drives the system to the all-$C$ state. For
large $b_0$ $(\geq 3.5)$, $\bar\rho$ goes toward its final value monotonically,
while, for smaller values of $b_0$, the convergence is
oscillatory. For $b_0 >b_{I,II}$, the system hits the fixed points
faster than the response from the environment can tune the
value of $b$. In an extended model where $D$ can appear, by
mutation, in an all-$C$ state, all-$C$ would not be evolutionary
stable.

In Fig.~\ref{rhocall}(d), we plot a diagram over the regions of $\alpha$-$b_0$ parameter space with distinct dynamic behavior.  We identify region I as when $\bar\rho$ at convergence is less than $0.5\%$ from $p^*$, i.e., $|\bar\rho-\rho^*|<0.005$ and region III as when the converged $\bar\rho$ is less than $0.005$. We note that the boundary value, $b_{\mathrm{I,II}}$, separating region I from II decreases with an increasing $\alpha$ ($b_{\mathrm{I,II}}\approx2$ for $\alpha\leq 3$ and $b_{\mathrm{I,II}}\rightarrow 1$ as $\alpha$ grows towards $11$).
In region I, for all measured values of $\alpha$, the system relaxes to a steady state with $\bar \rho\approx\rho^*$ and $\bar b$ converges to a stable value. For example, $b_0 =1.3$ gives $b(t\rightarrow\infty)\simeq 2.6$ [Figs.~\ref{rhocall}(b) and (c)]. This happens when the feedback in Eq.~(\ref{tupdate}) is strong enough to balance $b$. When $b_0$ increases beyond $b_{\mathrm{I,II}}$, the feedback from the environment starts affecting $b$ so strongly that the system inevitably hits an absorbing state. At a fixed point, $b$ grows  (if $\rho=1$) or decreases (if $\rho=0$) unboundedly. In this situation, as the fixed points in any real system would be metastable rather than permanent, $b$ should not be overinterpreted. Alternatively, one can limit the temptation by, in Eq.~(\ref{tupdate}), letting $b(t+1)=B$ if $b(t+1)>B$ and letting $b(t+1)=-B$ if $b(t+1)<-B$. If $B$ is large enough ($B\gtrsim 4$, for our parameter values). The conclusions from such a model are the same as for the one presented in this Letter, otherwise region II can vanish (results not shown). Preliminary studies suggest that an all-$C$ state also require a frequent updating of the strategies. Now strategies and links are updated equally often, but if the link update is 100 times more frequent than strategy updating, all-$C$ states almost never happen. If, on the other hand, the time scale is skewed in the other direction, the conclusions from Eq.~\ref{rhocall} remain the same. As a final note about Fig.~\ref{rhocall}(d), we see that $b_{\mathrm{II,III}}$, separating region II from III, increases monotonously with $\alpha$. That is, cooperation is enhanced by the feedback from the environment to the payoff matrices.

Now we turn to the connection between game dynamics and network structure. In this analysis, we only consider the network of long-range links, not the background square grid. In Fig.~\ref{rhocdk}, we show $\rho_k$, the fraction of cooperators or defectors of a particular degree $k$ in the steady state ($t>500$). The three different regions show different structure. For region I, represented by $b_0=1.3$ [Fig.~\ref{rhocdk}(a)], $\rho_k$ is larger for cooperators than defectors if $k\geq 3$. If $k\geq 77$, all nodes are cooperators. Since the final densities of $C$ and $D$ are equal in such situation, the high-degree $C$ can protect their neighbors from imitating defectors, and thus support cooperation. For region II, exemplified by $b_0=3.5$ [Fig.~\ref{rhocdk}(b)] where the steady state is all-$C$ (so $\rho_k=0$ for all $k$), we find that $\rho_k$ has a functional form closely described by a power-law with exponential cutoff and a decay exponent is about $2.7$. Since the steady state, in this case, is all-$C$, the payoff an agent can accrue will depend linearly on its degree. Consequently, during the process of rewiring, the probability of getting new links of the agents will be approximately proportional to the degrees they already have. In a strictly growing network, ``preferential attachment'' is known to generate a power-law degree distribution~\cite{ba:de}. In this case, with networks fixed in size, preferential attachment is not enough to explain the degree distribution. In such a case, the preferential attachment needs to be balanced by an antipreferential deletion of edges~\cite{salathe} in order for a power-law degree distribution to appear.  The power-law-like degree distribution remains for larger values of $b_0$ despite the different steady-state values of $\rho_k$. For a system in the all-$D$ state, the rewiring process behaves differently than in the all-$C$ case. Since the payoff a defector gets is independent of the total number of links it already has, its nonlocal link will be rewired randomly to another $D$, which generates networks with a Poisson degree distribution, as observed in Fig.~\ref{rhocdk}(c).

\begin{figure}
\includegraphics[width=\linewidth]{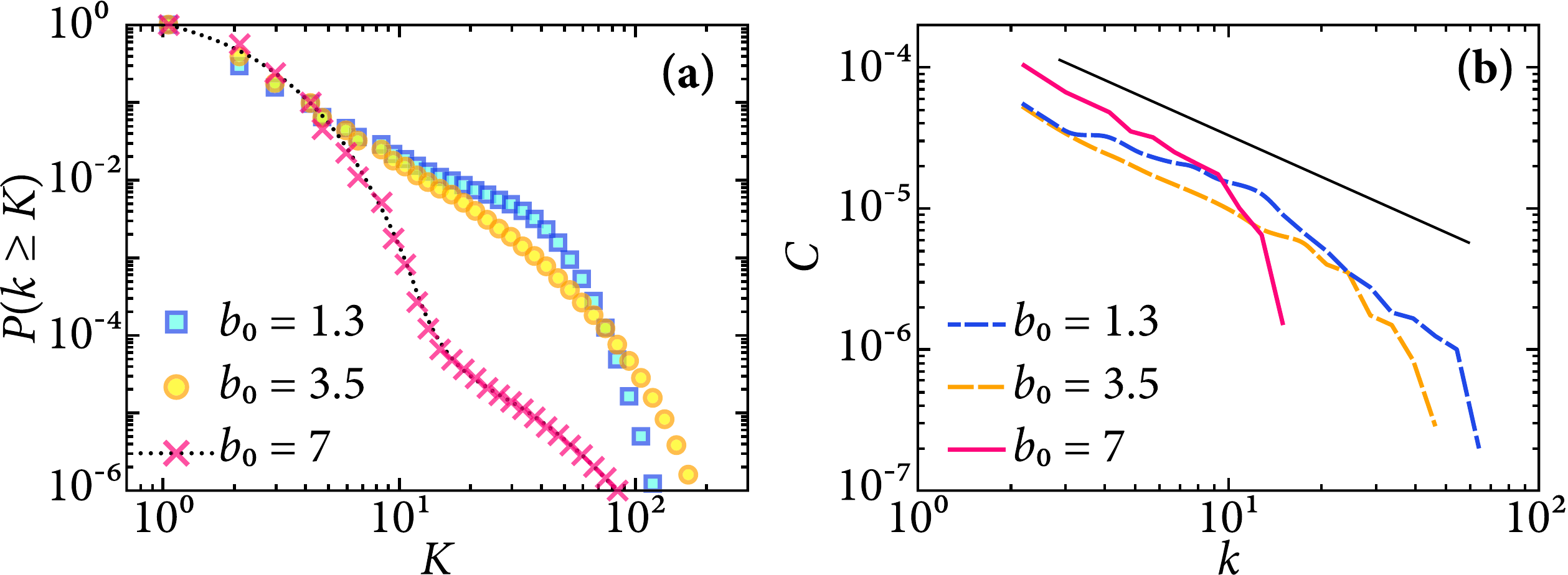}
\caption{(Color online) Structural properties of network in the steady state for different values of $b_0$. (a) displays the cumulative degree distribution.  The line for $b_0 =7$ follows a decay like a sum of an exponential and stretched exponential function.  (b) shows the clustering coefficient $C$ as a function of degree $k$. The line marks a scaling with exponent $-1$.}\label{netstruct}
\end{figure}

 Fig.~\ref{rhocdk} suggests that the coevolution of the contact patterns and the payoff matrix, in region II, makes the underlying network change from its initially random state to a heterogeneous structure. As shown in Fig.~\ref{netstruct}(a), the cumulative degree distribution $P(k\geq K)$ (the probability an observed degree $k$ is larger than $K$) depends strongly on $b_0$. Especially for $b_0 =3.5$ where, the distribution follows a power-law over two decades. For sufficiently large $b_0$, we observe a decay of the form $P \sim A\exp(-K/K_0)+B\exp(-K^{\tau}/K_1)$ ($K_{0,1}$ are fitting parameters)---a combination of an exponential and a stretched exponential form. The stretched exponential can, as mentioned above, be generated by a (non-linear) preferential attachment~\cite{krapi}. In Fig.~\ref{netstruct}(b), we investigate the hierarchical features of the steady state networks in greater detail. It has been argued that a characteristic feature of hierarchical networks is that the clustering coefficient (the fraction of possible triangles a node is member of with given the degree) is inversely proportional to degree~\cite{bara:modhie}. This is indeed what we observe for large $b_0$ values.

In conclusion, we have studied a game-theoretical model with feedback from the behavior of the agents to the rules of the game, via the payoff matrix, and an active optimization of both the contact structure between the agents and their strategies. With respect to the average cooperation density, the model is a non-equilibrium model. This makes the initial temptation value $b_0$ a crucial model parameter. We identify three regions of distinct dynamic behavior. In region I, the average cooperator density relaxes to a stable level through damped oscillations; in region III the systems reaches an all-defect state. For intermediate $b_0$-values (region II), the system ends at one of three fixed points, $0$, $\rho^*$ or $1$, with parameter-dependent probabilities. For some parameter values in this region, the system will almost certainly reach an all-Cooperator state. The all-cooperator state is absorbing, but if one extends this model to a non-equilibrium model, it would not be stable to mutations in $u$. In the all-$C$ state, the network has the most heterogeneous degree distribution, and also a clear $C\sim 1/k$ scaling of the clustering coefficient. Ref.~\cite{bara:modhie} argues that this feature is indicative of a hierarchical organization of the system. This is in contrast to usual explanations of social hierarchies as resulting from external factors such as age or fitness~\cite{wilson} or internal heterogeneities. The latter case is true also for our model in the limit of no environmental feedback, in which case it reduces to the model of Ref.~\cite{wli2007}.
But also the network dynamics is needed for the hierarchical topology and cooperation to co-emerge. If there is no network dynamics, the cooperation stabilizes at some intermediate $\rho$-value and does not reach the all-$C$ state.
In this case a power-law degree distribution emerges for intermediate cooperator levels. In other game-theoretic situations, hierarchical organization has sometimes proven to support cooperation~\cite{vukov:pd_santos:pd}, sometimes destabilizing it~\cite{fu:evoweak}. The source of the co-emergence of cooperation and a hierarchical topology in our model comes from the cooperators being stabilized by high-degree nodes, while there is no similar effect for the defectors. A similar positive feedback mechanisms between degree and payoff of cooperators drive the emergence of cooperation in the model of Ref.~\cite{zsch:2010}. This model differs from ours in that the payoff matrix is fixed and not a function of the state of the system.

In summary, our work shows a new possible mechanism for the coemergence of hierarchical structures and cooperation.  We foresee more studies of games in flexible settings where the game itself determines its rules and the player can choose when~\cite{zxw:prefpd} and with whom~\cite{egui:evo_ptn:coevo} to interact from its strategy.

\acknowledgments{This research was supported by the Wenner--Gren Foundations (S.L.), the Swedish Foundation for Strategic Research (P.H.), the Swedish Research Council (Z.X.W. and P.H.), and the WCU program through NRF Korea funded by MEST R31-2008-000-10029-0 (PH).}

\end{document}